\documentclass[letterpaper]{article} 
\usepackage[preprint]{aaai2027}  
\usepackage[hyphens]{url}  
\usepackage{graphicx} 
\usepackage{natbib}  
\usepackage{caption}
\usepackage{algorithm}
\usepackage{algorithmic}
\usepackage{amsmath}
\usepackage{amssymb}
\usepackage{booktabs}
\usepackage{multirow}
\usepackage{graphicx}
\usepackage{newfloat}
\usepackage{listings}
\DeclareCaptionStyle{ruled}{labelfont=normalfont, labelsep=colon, strut=off} 
\floatstyle{ruled}
\newfloat{listing}{tb}{lst}{}
\floatname{listing}{Listing}

\usepackage{booktabs}

\title{From Segments to Trajectories: Evolving Affective Graphs with Evidence Retrieval for Continuous EEG Emotion Recognition}
\author{
Chi Yang\textsuperscript{\rm 1}\equalcontrib,
Jihong Wang\textsuperscript{\rm 2}\equalcontrib,
Chengxi Xie\textsuperscript{\rm 3}\equalcontrib,
Kai He\textsuperscript{\rm 4},
Huan Liu\textsuperscript{\rm 2},
Man Yao\textsuperscript{\rm 5},
Shile Qi\textsuperscript{\rm 1}\corresponding,
Yuzhe Zhang\textsuperscript{\rm 1}\corresponding,
}

\affiliations{
\textsuperscript{\rm 1}College of Artificial Intelligence,
Nanjing University of Aeronautics and Astronautics\\
\textsuperscript{\rm 2}School of Computer Science and Technology,
Xi'an Jiaotong University\\
\textsuperscript{\rm 3}School of Intelligent Science and Engineering,
Harbin Institute of Technology (Shenzhen)\\
\textsuperscript{\rm 4}School of Public Health,
National University of Singapore\\
\textsuperscript{\rm 5}Institute of Automation,
Chinese Academy of Sciences\\

}

\begin{document}
\maketitle
\begin{abstract}
Electroencephalography (EEG)-based emotion recognition is important for affective computing and human-computer interaction, yet most existing methods divide a long trial into short segments and assign each segment the label of its source trial. Although this strategy increases the number of training samples, it reduces an evolving emotional response to a segment-level, coarse-grained, and static prediction problem. In reality, emotion may continuously emerge, intensify, weaken, and fluctuate as a stimulus unfolds, motivating the prediction of a time-aligned affective trajectory from the complete EEG trial. This task requires coordinated modeling of how spatial neural organization evolves throughout the trial and how local emotional fluctuations interact with longer-term trends. In this work, we formally define and systematically investigate continuous EEG emotion recognition as whole-trial affective trajectory prediction. We propose \textbf{EAGER}, an \textbf{E}volving \textbf{A}ffective \textbf{G}raph framework with \textbf{E}vidence \textbf{R}etrieval for continuous EEG emotion recognition. EAGER comprises two complementary modules: Affective State-guided Topology Evolution models the evolving spatial organization of EEG activity, while Multi-scale Temporal Evidence Retrieval integrates short-term fluctuations with longer-range temporal trends for time-aligned prediction. Experiments on MAHNOB-HCI, SEED-VII, and REFED show consistent gains in trajectory-tracking metrics over representative methods, with competitive pointwise errors.
\end{abstract}

\section{Introduction}

Emotion plays an important role in perception, decision making, and human behavior, positioning its automatic recognition as a fundamental challenge in affective computing and brain-computer interfaces ~\cite{picard1997affective,poria2017affective}. Among various affective cues, electroencephalography (EEG) offers non-invasive, temporally precise neural measurements, making it highly promising for affect-aware human-computer interaction and mental-state monitoring ~\cite{craik2019review,li2022eegreview}.

\begin{figure}[t]
\centering
\includegraphics[width=\columnwidth]{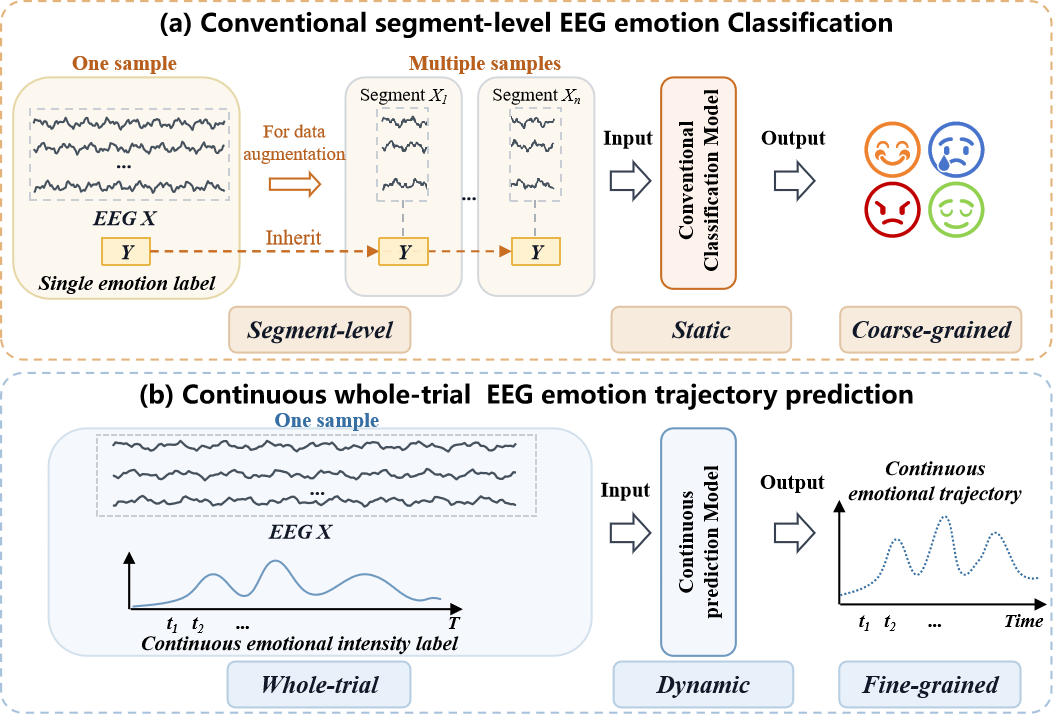}
\setlength{\abovecaptionskip}{3pt}
\caption{Comparison between conventional segment-level EEG emotion classification and continuous whole-trial EEG emotion trajectory prediction.}
\label{fig1}
\end{figure}

Most existing EEG emotion recognition studies follow the paradigm illustrated in Fig.~\ref{fig1}(a). Because controlled EEG acquisition and temporal emotion annotation are costly, participants are usually exposed to a limited number of relatively long emotional stimuli, and the entire EEG trial is assigned a single label. As each subject then contributes only tens of annotated trials, existing methods commonly divide every trial into short segments, often one second in length, and let each segment inherit the label of its source trial. Although this operation substantially increases the number of training samples, it reduces the task to \textit{segment-level}, \textit{static}, and \textit{coarse-grained} emotion classification.

However, emotional responses do not generally remain unchanged throughout a long stimulus. They may gradually emerge, intensify, weaken and fluctuate as the stimulus unfolds~\cite{russell1980circumplex,gaviria2021inertia}. As shown in Fig.~\ref{fig1}(b), a more natural paradigm is to process the whole EEG trial as a coherent sequence and predict a time-aligned affective trajectory. We refer to this task as continuous EEG emotion recognition, which shifts the modeling objective from isolated segment classification to \textit{whole-trial}, \textit{dynamic}, and \textit{fine-grained} emotion prediction. Besides describing when and how strongly an emotion changes, the resulting trajectory can help locate emotionally salient moments and support applications such as content-preference analysis, emotional-highlight detection and continuously responsive human-computer interaction.

The emergence of continuously annotated EEG datasets has provided an empirical basis for studying this paradigm. MAHNOB-HCI provides continuous affect annotations for video-induced emotional responses, SEED-VII records time-varying intensities of basic emotions, and REFED contains real-time valence and arousal trajectories synchronized with EEG recordings ~\cite{soleymani2012mahnob,jiang2025seedvii,ning2025refed}, yet most existing approaches simply extend segment-level pipelines by encoding local windows and aggregating them over time. Such a strategy misses the defining characteristic of the task: emotion unfolds as a coherent process in which spatial neural organization and temporal affective context evolve together. It also provides limited coordination between transient local fluctuations and slower trends spanning the complete trial.

In this work, we formally define and systematically investigate continuous EEG emotion recognition as whole-trial, time-aligned affective trajectory prediction. We propose \textbf{EAGER}, an \textbf{E}volving \textbf{A}ffective \textbf{G}raph framework with \textbf{E}vidence \textbf{R}etrieval that jointly models the evolving spatial organization of EEG activity and the multi-scale temporal evidence underlying continuous emotion changes. It consists of two complementary stages. First, \textit{Affective State-guided Topology Evolution (ASTE)} estimates short-term channel coupling from raw EEG, adaptively integrates it with mid- and long-range EEG contexts, and propagates a slowly varying graph state across successive windows, so that the resulting dynamic topology enhances the frequency-domain EEG representation at each time step. Second, \textit{Multi-scale Temporal Evidence Retrieval (MTER)} retrieves complementary evidence at multiple temporal resolutions from the topology-enhanced sequence, allowing each prediction to integrate local affective fluctuations with longer-term trial trends. By coordinating graph evolution with temporal evidence retrieval, EAGER represents an EEG trial as a continuously developing spatiotemporal process rather than a collection of independently processed segments. The contributions of this paper are summarized as follows:
\begin{itemize}
    \item We formally define and systematically study continuous EEG emotion recognition as whole-trial, time-aligned affective trajectory prediction. We distinguish this setting from the dominant segment-level static classification paradigm and evaluate it across three continuously annotated EEG datasets.
    \item We propose EAGER, a unified framework that combines evolving affective graph modeling with multi-scale temporal evidence retrieval. ASTE maintains a context-adaptive graph state across the complete trial, while MTER integrates local fluctuations and long-range trends for continuous emotion prediction.
    \item We conduct extensive experiments on MAHNOB-HCI, SEED-VII, and REFED. Comparisons with representative methods, detailed ablation studies, and qualitative trajectory analyses show the effectiveness of EAGER across diverse continuous emotion recognition settings.
\end{itemize}

\section{Related Work}
\subsection{Conventional EEG Emotion Recognition}
Most prior work formulates EEG-based emotion recognition as a static, segment-level classification task. Long-duration trials in benchmarks such as SEED and DEAP are assigned a single emotion category or global affective label~\cite{zheng2015seed,koelstra2012deap}, and are then partitioned into fixed-length windows, with every window inheriting the label of its source trial~\cite{chen2026emod}. Early studies combined handcrafted time- or frequency-domain features such as differential entropy with conventional classifiers~\cite{duan2013de}, after which convolutional, recurrent, and Transformer networks were adopted to learn hierarchical representations from segmented windows~\cite{song2020dgcnn,zhong2022rgnn,ding2023lggnet,philipchen2025adamgraph}. Graph-based models further exploit inter-channel dependencies~\cite{zhang2020vpr,song2021viag,zhang2021sparsedgcnn}, including causal graph formulations of emotional EEG~\cite{xiao2025dcgnn}. To handle inter-subject variability, domain adaptation and contrastive learning are used to improve generalization~\cite{zhao2021plug,wang2024dmmr,weng2026statemamba,liu2025multitosingle}. Yet these methods still process windows independently under static labels.

\subsection{Continuous EEG Emotion Recognition}
In contrast, continuous EEG approaches explicitly model temporal dependencies across successive windows. GIGN organizes short-window EEG graphs as nodes of a temporal graph to learn nested spatiotemporal dependencies~\cite{ding2023gign}; MASA-TCN adds space-aware temporal layers with attentive multi-anchor fusion~\cite{ding2024masa}; and EmT converts EEG into temporal graph sequences and applies a temporal contextual Transformer for cross-subject classification and regression~\cite{ding2025emt}. Cross-modal methods include Visual-to-EEG for video-to-EEG knowledge distillation~\cite{zhang2022visual},  MAET for flexible EEG and eye-movement inputs~\cite{jiang2023maet}, and TSMMF for bidirectional EEG--fNIRS fusion~\cite{si2025tsmmf}, with earlier studies also combining EEG and facial or visual cues for continuous emotion detection~\cite{soleymani2016continuous,choi2020multimodal}. However, these methods use either fixed connectivity or independently reconstructed window-level graphs, rather than maintaining an evolving graph state over the complete trial.

\begin{figure*}[t]
\centering
\includegraphics[width=0.9\textwidth]{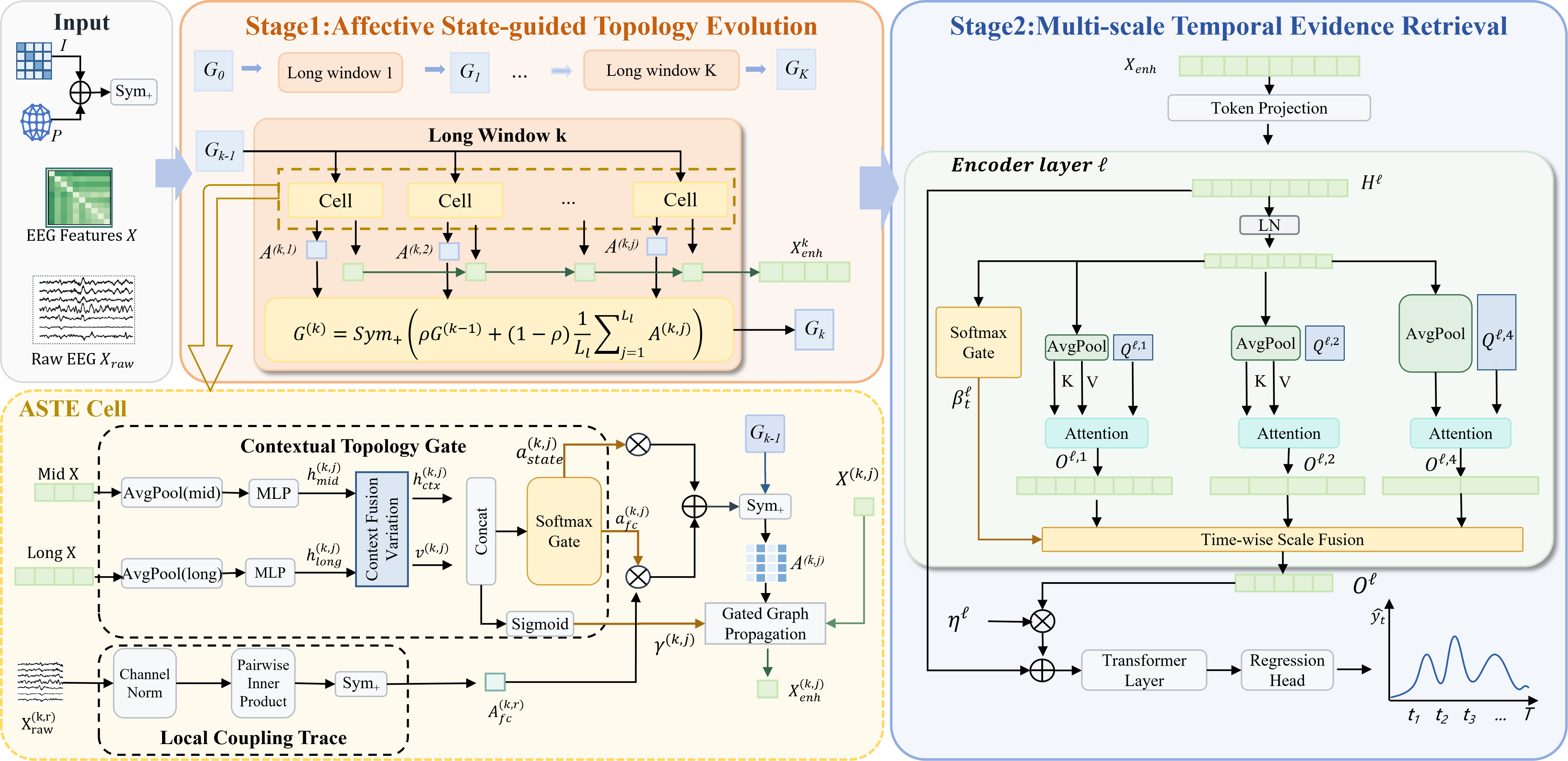} 
\setlength{\abovecaptionskip}{3pt}
\caption{Overall architecture of EAGER. Each ``Cell'' is the per-time-step unit of ASTE, combining the Local Coupling Trace and the Contextual Topology Gate to produce the dynamic graph $A^{(k,j)}$ and the enhanced feature $X^{(k,j)}_{\mathrm{enh}}$.}
\label{fig2}
\end{figure*}
\section{Method}
\subsection{Task Definition and Overall Framework}

\noindent\textbf{Task Definition.}
Conventional EEG emotion recognition typically adopts a static, segment-level paradigm (see Fig.~\ref{fig1}(a)), formalized as $\hat{y} = f(x_i)$, where an isolated short segment $x_i$ is mapped to a static trial-level label $y$. In contrast, we re-formulate continuous EEG emotion recognition as a \textit{whole-trial, time-aligned affective trajectory prediction} problem (see Fig.~\ref{fig1}(b)). Given a complete EEG trial, the objective is to capture the continuously unfolding emotional process. Let $X_{\mathrm{raw}} \in \mathbb{R}^{C \times T_{\mathrm{raw}}}$ denote the multi-channel raw EEG signals of the complete trial, where $C$ is the number of channels and $T_{\mathrm{raw}}$ is the number of sampling points. Let $X \in \mathbb{R}^{T \times C \times F}$ denote the corresponding frequency-domain feature sequence over $T$ time steps, where $F$ is the feature dimension per channel. Our framework learns a mapping $f$ to predict a continuous affective trajectory:
\begin{equation}
f(X_{\mathrm{raw}}, X) = \hat{Y}, \qquad \hat{Y} = \{\hat{y}_t\}_{t=1}^{T}, \quad \hat{y}_t \in \mathbb{R},
\end{equation}
where $\hat{Y}$ approximates the ground-truth emotional trajectory $Y = \{y_t\}_{t=1}^{T}$. A separate scalar regression model is trained for each affective dimension provided by the dataset.

\noindent\textbf{Overall Framework.}
As illustrated in Fig.~\ref{fig2}, the proposed EAGER, an \textbf{E}volving \textbf{A}ffective \textbf{G}raph framework with \textbf{E}vidence \textbf{R}etrieval, consists of two complementary stages: Affective State-guided Topology Evolution (ASTE) evolves a trial-level channel graph from raw EEG and enhances the frequency-domain features, while Multi-scale Temporal Evidence Retrieval (MTER) retrieves temporal evidence across scales from the enhanced sequence to produce the time-aligned predicted trajectory $\hat{Y}$.
\noindent\textbf{Notation.}
We partition the $T$ time steps into $K$ long windows of length $L_l$ ($T=K L_l$), each further divided into $R$ short windows of length $L_s$ ($L_l=R L_s$). Let indices $t$, $k$, $r$, and $j$ denote the global time step, long window, short window, and local time step, respectively, where $t = (k-1)L_l + j$. For the $k$-th long window, the frequency-domain feature at local step $j$ is $X^{(k, j)} \in \mathbb{R}^{C \times F}$, with its corresponding target $y^{(k, j)}$ and prediction $\hat{y}^{(k, j)}$. The raw EEG within the $r$-th short window (spanning local steps $j=(r-1)L_s+1, \dots, rL_s$) is denoted by $X_{\mathrm{raw}}^{(k, r)} \in \mathbb{R}^{C \times S_s}$, where $S_s$ is the number of raw samples.




\subsection{Affective State-guided Topology Evolution}

Short-term EEG functional connectivity is susceptible to local noise, whereas a fixed graph cannot represent changing connectivity patterns within a trial \cite{friston1994functional, hutchison2013dynamic, preti2017dynamic}. ASTE therefore comprises three successive components. Local Coupling Trace (LCT) estimates short-window connectivity from raw EEG. Contextual Topology Gate (CTG) generates time-step-level fusion conditions from mid- and long-range
contexts. Graph-State Evolution (GSE) propagates graph information between adjacent long windows. Together, these components generate $A^{(k, j)}$ and enhance $X^{(k, j)}$. The dynamic graph is a task-oriented representation of channel relationships and is not interpreted as causal neurophysiological connectivity.

\subsubsection{Graph-State Initialization}

At the beginning of each trial, ASTE initializes the graph state using an identity matrix $I$ and a fixed channel-structure prior $P$, which is a non-learnable electrode-distance graph built from a Gaussian kernel over channel coordinates.
\begin{equation}
G^{(0)}=\operatorname{Sym}_{+}(I+P), 
\label{eq:graph_initialization}
\end{equation}
where the non-negative symmetrization operator is
\begin{equation}
\operatorname{Sym}_{+}(M)=
\frac{1}{2}
\left[
\operatorname{ReLU}(M)
+
\operatorname{ReLU}(M)^\top
\right].
\label{eq:symmetrization}
\end{equation}
The prior $P$ is used only to initialize $G^{(0)}$ and is not repeatedly added to subsequent dynamic graphs.

\subsubsection{Short-Window Local Coupling Trace}
LCT estimates local channel connectivity from $X_{\mathrm{raw}}^{(k, r)}$. Let $\operatorname{Norm}_{\mathrm{ch}}(\cdot)$ denote channel-wise mean removal and $L_2$ normalization along the sampling dimension. The connectivity matrix is computed as
\begin{equation}
A_{\mathrm{fc}}^{(k, r)}
=
\operatorname{Sym}_{+}\!\left(
\operatorname{Norm}_{\mathrm{ch}}(X_{\mathrm{raw}}^{(k, r)})
\operatorname{Norm}_{\mathrm{ch}}(X_{\mathrm{raw}}^{(k, r)})^\top
\right).
\label{eq:local_connectivity}
\end{equation}
Here, $A_{\mathrm{fc}}^{(k, r)}\in\mathbb{R}^{C\times C}$ is shared by all $j\in S^{(k, r)}$. It provides short-window connectivity evidence.

\subsubsection{Contextual Topology Gate}

For each time step $j$ in the $k$-th long window, CTG extracts mid- and long-range contexts from $X^{(k)}$. Specifically, $h_{\mathrm{mid}}^{(k, j)}$ and $h_{\mathrm{long}}^{(k, j)}$ are obtained using sliding average pooling with lengths $L_m$ and $L_l$, respectively, followed
by learnable mappings. The change in long-range context is defined as $\Delta h^{(k, j)} =h_{\mathrm{long}}^{(k, j)}-h_{\mathrm{long}}^{(k, j-1)}$ and is set to zero when $j=1$.

CTG computes the fused context and contextual change intensity as
\begin{equation}
\begin{aligned}
h_{\mathrm{ctx}}^{(k, j)}
&=
\phi_{\mathrm{ctx}}\!\left([h_{\mathrm{mid}}^{(k, j)};h_{\mathrm{long}}^{(k, j)};
 \Delta h^{(k, j)}]
\right), \\
v^{(k, j)}
&=\sigma\!\left(\phi_v\!\left([h_{\mathrm{mid}}^{(k, j)};h_{\mathrm{long}}^{(k, j)};|\Delta h^{(k, j)}|]
\right)
\right).
\end{aligned}
\label{eq:context_fusion}
\end{equation}
Here, $h_{\mathrm{ctx}}^{(k, j)}$ denotes the fused context, and $v^{(k, j)}\in(0, 1)$ denotes contextual change intensity.

The fusion weights for the historical graph state and the current local connectivity are generated by
\begin{equation}
[\alpha_{\mathrm{state}}^{(k, j)}, 
 \alpha_{\mathrm{fc}}^{(k, j)}]
=
\operatorname{Softmax}\!\left(
g_\alpha([h_{\mathrm{ctx}}^{(k, j)};v^{(k, j)}])
\right), 
\label{eq:graph_gate}
\end{equation}
where
$\alpha_{\mathrm{state}}^{(k, j)}
+\alpha_{\mathrm{fc}}^{(k, j)}=1$.
Although $A_{\mathrm{fc}}^{(k, r)}$ remains fixed within the same short window, the fusion weights are independently generated at each time step.

\subsubsection{Graph-State Evolution and Feature Enhancement}
For each $j\in S^{(k, r)}$, CTG combines the local connectivity with the graph state propagated from the preceding long window:
\begin{equation}
A^{(k, j)}
=
\operatorname{Sym}_{+}\!\left(
\alpha_{\mathrm{state}}^{(k, j)}G^{(k-1)}
+
\alpha_{\mathrm{fc}}^{(k, j)}A_{\mathrm{fc}}^{(k, r)}
\right).
\label{eq:dynamic_graph}
\end{equation}

After processing the $k$-th long window, GSE updates the graph state by aggregating its dynamic graphs:
\begin{equation}
G^{(k)}
=
\operatorname{Sym}_{+}\!\left(
\rho G^{(k-1)}
+
(1-\rho)\frac{1}{L_l}
\sum_{j=1}^{L_l}A^{(k, j)}
\right), 
\label{eq:graph_state}
\end{equation}
where $\rho\in[0, 1]$ controls graph-state retention across adjacent long windows.

Following the neighborhood-aggregation principle of GNNs \cite{defferrard2016chebnet, kipf2017gcn}, CTG generates a feature-wise gate that controls the injection of graph-propagated information:
\begin{equation}
\begin{aligned}
\gamma^{(k, j)}
&=
\sigma\!\left(
g_\gamma([h_{\mathrm{ctx}}^{(k, j)};v^{(k, j)}])
\right), \\
X_{\mathrm{enh}}^{(k, j)}
&=
(1-\gamma^{(k, j)})\odot X^{(k, j)}\\
&\quad+
\gamma^{(k, j)}\odot
\left(A^{(k, j)}X^{(k, j)}\right).
\end{aligned}
\label{eq:feature_enhancement}
\end{equation}
Here, $\gamma^{(k, j)}\in(0, 1)^F$ is broadcast along the channel dimension, gating how much dynamic-graph information is fused into the original frequency-domain features.

\begin{table*}[t]
\centering
\renewcommand{\arraystretch}{0.9}
\setlength{\tabcolsep}{3.0pt}
\renewcommand{\arraystretch}{1.0}
\resizebox{\textwidth}{!}{
\begin{tabular}{lcccccccc}
\toprule
\multirow{2}{*}{Method}
& \multicolumn{4}{c}{MAHNOB-HCI}
& \multicolumn{4}{c}{SEED-VII} \\
\cmidrule(lr){2-5} \cmidrule(lr){6-9}
& CCC$\uparrow$ & PCC$\uparrow$ & MAE$\downarrow$ & RMSE$\downarrow$
& CCC$\uparrow$ & PCC$\uparrow$ & MAE$\downarrow$ & RMSE$\downarrow$ \\
\midrule
SVR
& 34.38$\pm$22.35 & 44.19$\pm$21.95 & 11.19$\pm$3.54 & 14.08$\pm$3.97
& 2.98$\pm$4.96 & 4.13$\pm$7.47 & 28.77$\pm$3.74 & 34.97$\pm$3.84 \\
KNN
& 18.87$\pm$14.28 & 26.51$\pm$18.15 & 10.75$\pm$3.91 & 12.12$\pm$3.63
& 1.01$\pm$2.85 & 1.24$\pm$3.84 & 29.01$\pm$3.58 & 34.61$\pm$3.81 \\
LSTM
& 30.27$\pm$21.10 & 36.73$\pm$24.91 & 6.44$\pm$1.86 & 8.38$\pm$2.58
& 15.97$\pm$9.73 & 18.76$\pm$9.73 & 30.62$\pm$5.46 & 37.07$\pm$5.84 \\
Visual-to-EEG
& 39.15$\pm$23.25 & 50.36$\pm$23.37 & 5.55$\pm$2.85 & 6.91$\pm$2.63
& 17.89$\pm$7.96 & 22.32$\pm$7.88 & 30.43$\pm$4.45 & 37.03$\pm$4.85 \\
GIGN
& 30.25$\pm$25.16 & 42.77$\pm$29.56 & 10.08$\pm$4.14 & 13.12$\pm$4.95
& 7.67$\pm$6.23 & 9.91$\pm$8.04 & 31.38$\pm$3.04 & 37.84$\pm$3.33 \\
MAET
& 35.22$\pm$22.58 & 41.56$\pm$22.52 & 5.14$\pm$2.01 & 6.88$\pm$2.58
& 4.22$\pm$6.81 & 5.18$\pm$8.72 & 33.27$\pm$5.24 & 40.27$\pm$5.80 \\
MASA-TCN
& \underline{40.69$\pm$24.16} & 51.89$\pm$24.17 & 5.29$\pm$2.03 & 6.87$\pm$2.32
& \underline{30.36$\pm$8.16} & \underline{35.65$\pm$6.25} & \underline{27.94$\pm$5.17} & \underline{34.02$\pm$5.46} \\
TSMMF
& 19.23$\pm$19.52 & 38.58$\pm$24.40 & \textbf{4.23$\pm$1.40} & \textbf{5.93$\pm$1.78}
& 11.01$\pm$8.53 & 13.95$\pm$8.92 & 34.21$\pm$6.87 & 41.30$\pm$7.25 \\
EmT
& 40.15$\pm$23.78 & 52.79$\pm$23.63 & 5.70$\pm$2.48 & 7.21$\pm$2.80
& 13.20$\pm$5.82 & 16.53$\pm$5.50 & 32.98$\pm$6.38 & 39.92$\pm$6.58 \\
Ours
& \textbf{44.43$\pm$23.10} & \textbf{55.42$\pm$21.32} & \underline{5.08$\pm$2.32} & \underline{6.62$\pm$2.63}
& \textbf{51.13$\pm$11.29} & \textbf{59.45$\pm$10.47} & \textbf{25.44$\pm$4.73} & \textbf{31.45$\pm$6.76} \\
\bottomrule
\end{tabular}
}
\caption{ Experimental results on MAHNOB-HCI and SEED-VII under the leave-one-subject-out (LOSO) protocol.}
\label{tab:main_results}
\end{table*}

\subsection{Multi-scale Temporal Evidence Retrieval}
The ASTE output $X_{\mathrm{enh}}^{(k, j)}$ encodes time-varying spatial topology. MTER subsequently models temporal dependencies over the complete
trial without updating the EEG graph. It instead performs multi-scale contextual retrieval and temporal refinement on the topology-enhanced representations.

The ASTE outputs are arranged as a temporally ordered sequence $\{X_{\mathrm{enh}}^{(t)}\}_{t=1}^{T}$, where $T=KL_l$ and $t=(k-1)L_l+j$. MTER uses multi-head attention \cite{vaswani2017attention} with the temporal scale set $\mathcal{S}_{\mathrm{MTER}}=\{1, 2, 4\}$. Inspired by the Temporal Query mechanism \cite{lin2025tqnet}, it adapts learnable queries to time-aligned continuous EEG regression through multi-scale pooling and adaptive fusion.

Each $X_{\mathrm{enh}}^{(t)}$ is vectorized and mapped by a learnable projection into a $D$-dimensional token, $z_t=\operatorname{Proj}(\operatorname{vec}(X_{\mathrm{enh}}^{(t)}))
\in\mathbb{R}^{D}$. The resulting sequence is $Z=[z_1, \ldots, z_T]^\top\in\mathbb{R}^{T\times D}$.

Let $H^{(0)}=Z$, and let $U^{(\ell)}=\operatorname{LN}(H^{(\ell)})$ denote the normalized input of the $\ell$-th encoding layer. For each scale $s$, MTER applies temporal average pooling to $U^{(\ell)}$ and uses a learnable query $Q^{(\ell, s)}$ aligned with the original time axis to retrieve temporal evidence:
\begin{equation}
\begin{aligned}
O^{(\ell, s)}
&=
\operatorname{MHA}\!\Bigl(
Q^{(\ell, s)}, \\
&\qquad
\operatorname{AvgPool}_s(U^{(\ell)}), 
\operatorname{AvgPool}_s(U^{(\ell)})
\Bigr).
\end{aligned}
\label{eq:mter_query}
\end{equation}

A time-step-specific gate fuses the retrieved evidence:
\begin{equation}
\begin{aligned}
\boldsymbol{\beta}_t^{(\ell)}
&=
\operatorname{Softmax}(g_\beta(U_{t, :}^{(\ell)})), \\
o_t^{(\ell)}
&=
\sum_{s\in\mathcal{S}_{\mathrm{MTER}}}
\beta_{t, s}^{(\ell)}[O^{(\ell, s)}]_{t, :}.
\end{aligned}
\label{eq:mter_fusion}
\end{equation}
where $\boldsymbol{\beta}_t^{(\ell)} =[\beta_{t,s}^{(\ell)}]_{s\in\mathcal{S}_{\mathrm{MTER}}}$
is the scale-weight vector at time step $t$.

Let $O^{(\ell)}=[o_1^{(\ell)}, \ldots, o_T^{(\ell)}]^\top$. The fused evidence is injected into the temporal refinement layer using a learnable strength $\eta^{(\ell)}$:
\begin{equation}
H^{(\ell+1)}
=
\mathcal{R}_{\ell}
\left(H^{(\ell)}+\eta^{(\ell)}O^{(\ell)}\right), 
\label{eq:mter_update}
\end{equation}
where $\mathcal{R}_{\ell}(\cdot)$ consists of normalization, temporal self-attention, feed-forward mapping, and residual connections. $\eta^{(\ell)}$ is a learnable scalar, not the output of a gating network. After $N_{\mathrm{enc}}$ encoding layers, the learnable regression head $f_{\mathrm{reg}}(\cdot)$ predicts
\begin{equation}
\hat y_t
=
f_{\mathrm{reg}}
\left([H^{(N_{\mathrm{enc}})}]_{t, :}\right), 
\label{eq:prediction}
\end{equation}

\subsection{Training Objective and Inference}
Let $Y=\{y_t\}_{t=1}^{T}$ and $\widehat Y=\{\hat y_t\}_{t=1}^{T}$ denote the ground-truth and predicted trajectories. The primary objective combines the concordance correlation coefficient (CCC)
\cite{lin1989ccc} and MSE losses:
\begin{equation}
\mathcal{L}_{\mathrm{main}}
=
\lambda_{\mathrm{ccc}}
\left[1-\operatorname{CCC}(\widehat Y, Y)\right]
+
\lambda_{\mathrm{mse}}
\operatorname{MSE}(\widehat Y, Y).
\label{eq:main_objective}
\end{equation}
In addition, several auxiliary terms supervise the mid- and long-range context branches, the contextual change intensity, and the diversity of the multi-scale queries; their definitions and weights are provided in the supplementary material.

\begin{table*}[t]
\centering
\setlength{\tabcolsep}{3.0pt}
\renewcommand{\arraystretch}{1.0}
\resizebox{\textwidth}{!}{
\begin{tabular}{lcccccccc}
\toprule
\multirow{2}{*}{Method}
& \multicolumn{4}{c}{REFED-Valence}
& \multicolumn{4}{c}{REFED-Arousal} \\
\cmidrule(lr){2-5} \cmidrule(lr){6-9}
& CCC$\uparrow$ & PCC$\uparrow$ & MAE$\downarrow$ & RMSE$\downarrow$
& CCC$\uparrow$ & PCC$\uparrow$ & MAE$\downarrow$ & RMSE$\downarrow$ \\
\midrule
SVR
& 12.74$\pm$23.24 & 17.36$\pm$30.31 & 20.92$\pm$5.63 & 26.18$\pm$6.46
& 13.31$\pm$16.27 & 19.96$\pm$24.06 & 17.89$\pm$5.44 & 22.45$\pm$6.05 \\
KNN
& 9.37$\pm$17.67 & 11.66$\pm$20.01 & 20.73$\pm$5.34 & 25.54$\pm$6.12
& 9.71$\pm$13.70 & 12.86$\pm$17.62 & 17.72$\pm$5.57 & 21.65$\pm$6.18 \\
LSTM
& 12.98$\pm$23.03 & 15.40$\pm$26.40 & 20.76$\pm$5.54 & 25.69$\pm$6.51
& 14.05$\pm$16.38 & 17.67$\pm$20.77 & 17.84$\pm$5.44 & 21.80$\pm$6.12 \\
Visual-to-EEG
& 13.98$\pm$25.76 & 15.61$\pm$29.12 & 22.18$\pm$6.53 & 27.23$\pm$7.56
& 14.81$\pm$18.68 & 17.32$\pm$21.37 & 18.91$\pm$5.63 & 23.22$\pm$6.34 \\
GIGN
& 12.43$\pm$24.98 & 14.78$\pm$28.79 & 21.44$\pm$6.37 & 26.39$\pm$7.37
& 14.12$\pm$16.93 & 17.28$\pm$20.97 & 18.25$\pm$5.55 & 22.36$\pm$6.21 \\
MAET
& 9.75$\pm$19.40 & 12.81$\pm$24.04 & \underline{20.63$\pm$5.54} & \underline{25.51$\pm$6.49}
& 9.99$\pm$14.62 & 14.55$\pm$20.20 & \underline{17.58$\pm$5.52} & \underline{21.42$\pm$6.20} \\
MASA-TCN
& 14.69$\pm$26.20 & 16.91$\pm$30.70 & 21.36$\pm$6.48 & 26.28$\pm$7.54
& 14.17$\pm$21.10 & 18.46$\pm$25.67 & 18.34$\pm$6.08 & 22.54$\pm$7.16 \\
TSMMF
& 13.67$\pm$19.95 & 15.12$\pm$22.02 & 23.06$\pm$5.56 & 28.64$\pm$6.78
& 14.23$\pm$16.36 & 16.21$\pm$18.45 & 19.78$\pm$5.75 & 24.35$\pm$6.60 \\
EmT
& \underline{15.22$\pm$26.86} & \underline{17.39$\pm$30.98} & 21.25$\pm$6.56 & 26.16$\pm$7.53
& \underline{16.69$\pm$19.72} & \underline{20.52$\pm$23.73} & 18.21$\pm$5.94 & 22.25$\pm$6.75 \\
Ours
& \textbf{20.42$\pm$29.37} & \textbf{24.71$\pm$33.13} & \textbf{20.26$\pm$6.40} & \textbf{24.98$\pm$7.54}
& \textbf{18.20$\pm$21.22} & \textbf{21.87$\pm$24.82} & \textbf{16.58$\pm$4.50} & \textbf{19.94$\pm$4.79} \\
\bottomrule
\end{tabular}
}
\caption{Experimental results on REFED (valence and arousal) under the subject-dependent 3-fold protocol.}
\label{tab:refed_results}
\end{table*}

\begin{figure*}[!t]
\centering
\includegraphics[width=1.0 \textwidth]{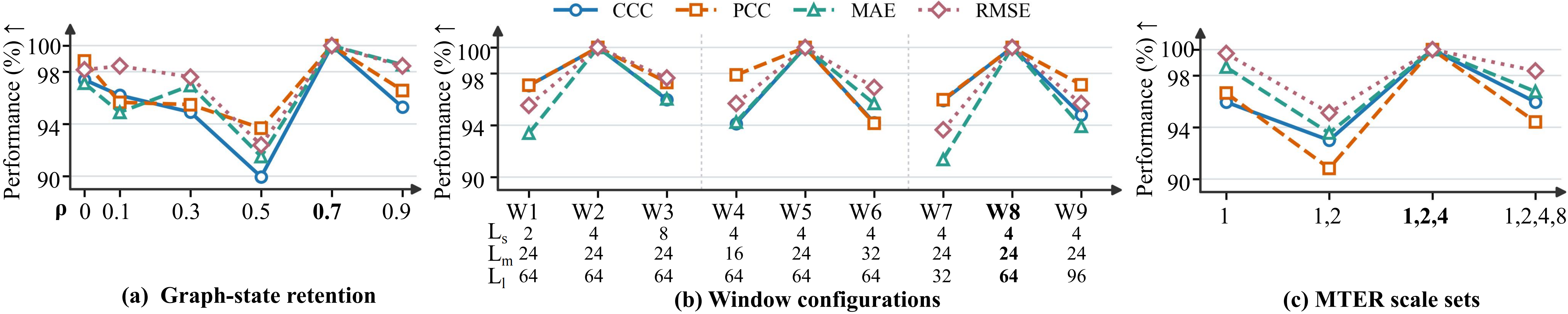} 
\setlength{\abovecaptionskip}{0pt}
\caption{The results of the sensitivity analysis. All metrics are normalized to the default configuration (= 100\%) on MAHNOB-HCI. The default configuration adopted in all other experiments is $\rho = 0.7$ in (a), $(L_s, L_m, L_l) = (4, 24, 64)$ (i.e., W8) in (b), and the scale set $\{1, 2, 4\}$ in (c).}
\label{fig3}
\end{figure*}

\section{Experiments}
This section presents the core experimental settings, main results, and analyses of EAGER. More detailed implementation settings, additional quantitative results, and statistical analyses are provided in the supplementary material. We evaluate EAGER on three video-elicited EEG emotion datasets, assessing its predictive performance, sensitivity to temporal configurations, the contribution of individual components, and qualitative trajectory analysis.
\subsection{Experiment Setup}
\paragraph{Datasets and protocols}
We evaluate EAGER on three video-elicited EEG datasets with time-indexed regression targets. These datasets are among the few publicly available ones that provide continuous affective trajectories consistent with our regression setting. MAHNOB-HCI~\cite{soleymani2012mahnob} and SEED-VII~\cite{jiang2025seedvii} target continuous valence and emotion-intensity trajectories, respectively: MAHNOB-HCI valence labels lie in $[-1,1]$, whereas the original SEED-VII intensity scores in $[0,100]$ are rescaled to $[0,1]$. REFED~\cite{ning2025refed} provides valence and arousal trajectories, whose original coordinates in $[1,255]$ are normalized to $[0,1]$. For evaluation, MAHNOB-HCI and SEED-VII use leave-one-subject-out (LOSO) validation, whereas REFED follows subject-dependent 3-fold validation. Each complete trial is the basic prediction unit, and predicted and reference sequences are concatenated within each test partition before scoring. Each reported result is averaged over 24 LOSO runs on MAHNOB-HCI, 20 on SEED-VII, and 96 subject-fold runs on REFED (32 subjects $\times$ 3 folds). We report four widely used and complementary metrics: the concordance correlation coefficient (CCC), Pearson correlation coefficient (PCC), mean absolute error (MAE), and root mean square error (RMSE), where CCC and PCC measure trajectory agreement while MAE and RMSE quantify pointwise prediction error. In all tables these metrics are multiplied by 100, with higher CCC/PCC and lower MAE/RMSE indicating better performance.
\paragraph{Implementation and baselines}
EAGER uses band-power features for MAHNOB-HCI and differential-entropy features~\cite{duan2013de} for SEED-VII and REFED, with a Transformer temporal backbone trained using AdamW~\cite{loshchilov2019adamw}. Hyperparameters are selected according to validation CCC, with key sensitivity ranges, final settings, and corresponding analysis reported in the Sensitivity Analysis section and supplementary material. We compare EAGER with nine baselines: SVR~\cite{smola2004svr}, KNN~\cite{cover1967nearest}, LSTM~\cite{hochreiter1997lstm}, Visual-to-EEG~\cite{zhang2022visual}, GIGN~\cite{ding2023gign}, MAET~\cite{jiang2023maet}, MASA-TCN~\cite{ding2024masa}, TSMMF~\cite{si2025tsmmf}, and EmT~\cite{ding2025emt}. The random number seed is set to 42.

\begin{figure*}[t]
\centering
\includegraphics[width=1.0\textwidth]{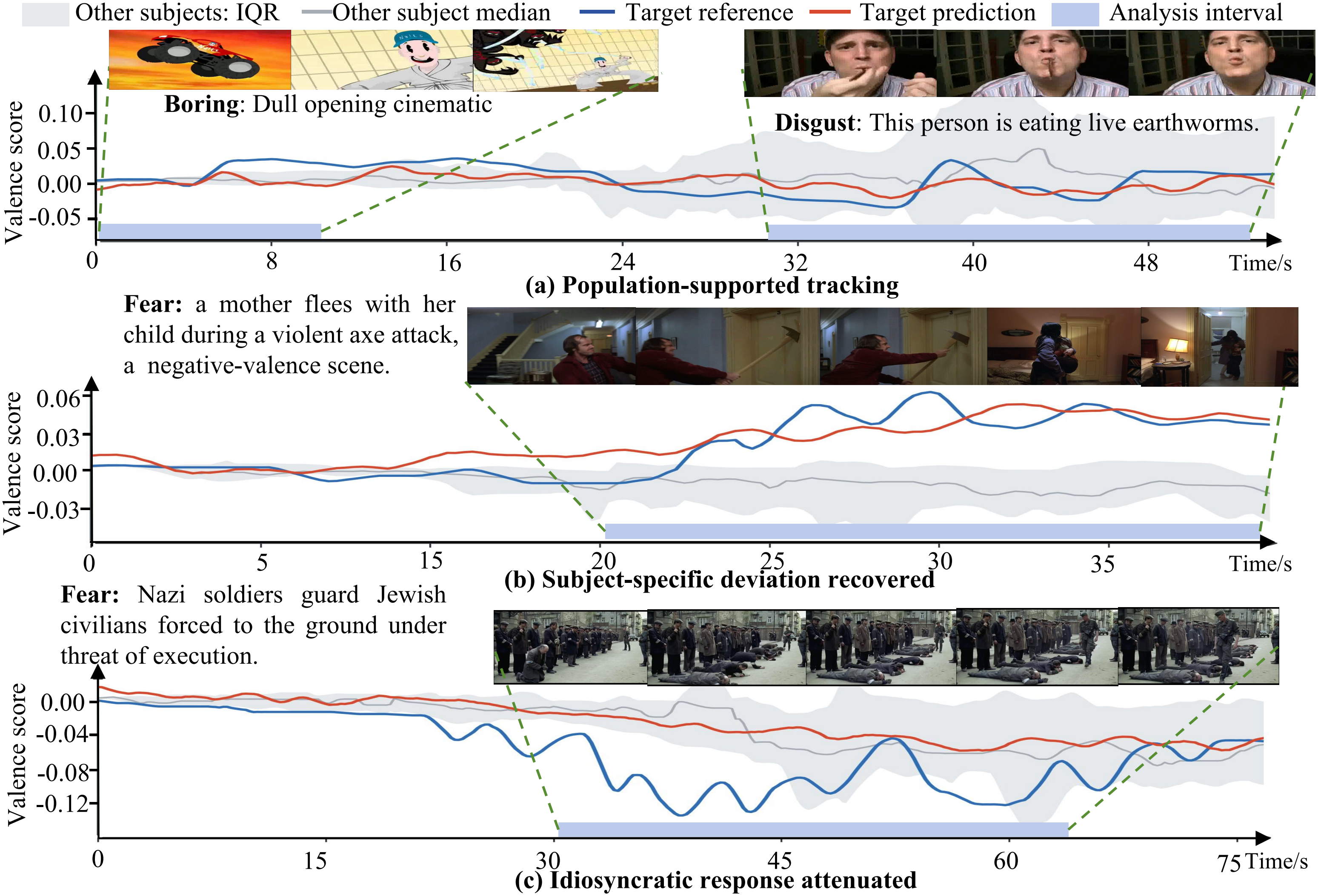} 
\captionsetup{skip=3pt}
\caption{Representative continuous predictions on MAHNOB-HCI.}
\label{fig4}
\end{figure*}
\subsection{Comparison Experiments}
We compare EAGER against nine baselines on MAHNOB-HCI, SEED-VII, and REFED under their respective evaluation protocols. Tables~\ref{tab:main_results} and~\ref{tab:refed_results} report the results, from which we draw the following three observations.

\noindent\textbf{Observation 1: EAGER achieves the best trajectory-tracking performance (CCC/PCC) across all datasets and stays competitive on MAE/RMSE.} These gains show that jointly modeling an evolving topology with multi-scale temporal evidence tracks continuous emotion more faithfully than static-graph or segment-aggregation baselines. The only exception is TSMMF, which attains lower MAE/RMSE but far weaker CCC/PCC ($19.23$/$38.58$), suggesting that it fits the average amplitude rather than the temporal variation; EAGER instead prioritizes temporal tracking while keeping the second-lowest MAE/RMSE.

\noindent\textbf{Observation 2: Methods that jointly model spatial and temporal structure perform best across all three datasets.} Among the baselines, spatiotemporal models such as MASA-TCN and EmT consistently rank highest, while methods that rely on spatial structure alone (GIGN) or temporal modeling alone (LSTM) are less stable, and static regressors (SVR, KNN) perform worst, nearly failing on SEED-VII (CCC below $3$). This pattern holds across the three datasets and indicates that spatial connectivity and temporal context are complementary. EAGER follows the same principle by coupling an evolving topology with multi-scale temporal retrieval, which explains its consistent advantage.

\noindent\textbf{Observation 3: All methods score low on REFED, whose stepwise labels let simple regressors appear competitive on pointwise metrics.} REFED performance is uniformly low (CCC below $21$). Its valence and arousal labels change stepwise and stay flat over long intervals, so predictions near the running mean already yield small MAE/RMSE; consequently SVR and KNN match several deep models on these metrics despite far lower CCC/PCC (e.g., SVR: $20.92$ MAE but $12.74$ CCC on REFED-Valence). CCC/PCC thus better reflect trajectory tracking, and EAGER achieves the highest CCC on both REFED dimensions.

To test whether the improvements hold across subjects, we run paired Wilcoxon signed-rank tests on subject-level CCC and PCC scores between EAGER and each deep-learning baseline on MAHNOB-HCI and SEED-VII under LOSO. Most comparisons give $p<0.05$, indicating that the gains are generally consistent at the subject level. Full results appear in the supplementary material.
\subsection{Sensitivity Analysis}
We conduct a sensitivity analysis on three key hyperparameters: graph-state retention $\rho$, temporal window lengths $(L_s,L_m,L_l)$, and the MTER scale set. All metrics in Fig.~\ref{fig3} are normalized to the adopted configuration on MAHNOB-HCI. As shown in Fig.~\ref{fig3}(a), performance varies non-monotonically with $\rho$, and $\rho=0.7$ provides the best overall result. Fig.~\ref{fig3}(b) shows that $(L_s,L_m,L_l)=(4,24,64)$ achieves the most balanced performance, whereas shorter or longer contexts degrade at least one metric. In Fig.~\ref{fig3}(c), the scale set $\{1,2,4\}$ outperforms both the single-scale and two-scale alternatives, while adding scale 8 provides no additional benefit. We therefore adopt $\rho=0.7$, $(4,24,64)$, and $\{1,2,4\}$ in subsequent experiments.

\begin{table}[t]
\centering
\setlength{\tabcolsep}{2.2pt}
\renewcommand{\arraystretch}{1.05}
\resizebox{\columnwidth}{!}{
\begin{tabular}{lcccccccc}
\toprule
\multirow{2}{*}{Method}
& \multicolumn{4}{c}{MAHNOB-HCI}
& \multicolumn{4}{c}{SEED-VII} \\
\cmidrule(lr){2-5} \cmidrule(lr){6-9}
& CCC$\uparrow$ & PCC$\uparrow$ & MAE$\downarrow$ & RMSE$\downarrow$
& CCC$\uparrow$ & PCC$\uparrow$ & MAE$\downarrow$ & RMSE$\downarrow$ \\
\midrule
LSTM
& 30.27 & 36.73 & 6.44 & 8.38
& 15.97 & 18.76 & 30.62 & 37.07 \\

w/o ASTE
& 39.75 & 51.27 & 5.96 & 7.53
& 46.31 & 56.23 & 27.47 & 33.76 \\

w/o CTG
& 41.30 & 51.49 & 5.54 & 7.05
& 48.27 & 57.40 & 26.63 & 32.80 \\

w/o GSE
& 40.09 & 51.20 & 5.40 & 6.92
& 48.82 & 57.73 & 26.31 & 32.42 \\

w/o MTER
& 37.31 & 44.79 & 5.25 & 6.84
& 47.12 & 54.83 & 26.14 & 32.12 \\

Full
& \textbf{44.43} & \textbf{55.42} & \textbf{5.08} & \textbf{6.62}
& \textbf{51.13} & \textbf{59.45} & \textbf{25.44} & \textbf{31.45} \\

\bottomrule
\end{tabular}
}
\caption{Ablation results on MAHNOB-HCI and SEED-VII.}
\label{tab:ablation}  
\end{table}

\subsection{Ablation Study}
We conduct ablation studies to verify the effectiveness of each component, evaluated on MAHNOB-HCI and SEED-VII under the LOSO protocol. We compare the full model against several variants: replacing the entire backbone with a plain LSTM, removing the ASTE module, ablating its CTG and GSE components individually, and removing the MTER module. LCT is not ablated alone, as it produces the connectivity that CTG and GSE rely on. All variants share the same training configuration and differ only in the ablated part, so the changes can be attributed to the corresponding component. The results are reported in Table~\ref{tab:ablation}.

From these results, we draw the following observations. (1) ASTE contributes the most, as removing it causes the largest degradation across metrics, confirming the value of adaptive graph evolution. (2) CTG and GSE are complementary, since ablating either one degrades all four metrics. (3) Multi-scale temporal retrieval is essential, and replacing it with a plain LSTM leads to the weakest performance overall. (4) The full model performs best on every metric, indicating that the spatial and temporal designs are mutually reinforcing.

\subsection{Visualization Analysis}
To compare EAGER's predictions with individual and population-level responses, Fig.~\ref{fig4} visualizes three representative MAHNOB-HCI valence trials. The blue and red curves denote the subject-specific ground truth and EAGER prediction, respectively. The grey band and line show the interquartile range and median of the remaining subjects' labels for the same stimulus. Thumbnails and shaded intervals indicate annotated stimulus events. Together, the three trials illustrate distinct model behaviors. Additional visualization examples are provided in the supplementary material.

In Fig.~\ref{fig4}(a), the reference, the prediction, and the population tendency stay close throughout the trial. EAGER tracks salient stimulus events, producing a mild positive response to the dull opening cinematic and a synchronized negative response to the disgusting clip, which shows that it captures event-driven affective changes rather than a static baseline.

Fig.~\ref{fig4}(b) shows a subject whose reference departs from the population: the fear-inducing scene elicits a positive rather than an aversive response. Here EAGER follows the subject's own trajectory instead of the group median, indicating that it recovers individualized responses rather than regressing toward the population mean.

Fig.~\ref{fig4}(c) presents the opposite situation. The subject's reference deviates strongly from the population, dropping well below the grey band, whereas the prediction remains closer to the population tendency and underestimates the magnitude of this deviation. This case indicates that EAGER tracks the direction of strong idiosyncratic responses but attenuates their extent, which we regard as a current limitation.

\section{Conclusion}
In this work, we formulate continuous EEG emotion recognition as whole-trial, time-aligned affective trajectory prediction, departing from the conventional segment-level classification paradigm. We propose \textbf{EAGER}, which couples Affective State-guided Topology Evolution, modeling the evolving spatial organization of EEG activity, with Multi-scale Temporal Evidence Retrieval, integrating local fluctuations and longer-term trends. Experiments on MAHNOB-HCI, SEED-VII, and REFED show that EAGER consistently improves trajectory-tracking metrics over representative baselines, and the trajectory-level analysis indicates that it tracks event-driven and subject-specific affective changes. Future work will extend EAGER toward causal, streaming inference for real-time affective interfaces, and toward better capturing extreme subject-specific responses.
\bibliography{aaai2027}
\end{document}